\documentclass[preprint]{aastex}
\journalid{submitted to Ap.J. }{ }
\begin{document} 


\title{A Helium White Dwarf of Extremely Low Mass } 

\author{James Liebert\altaffilmark{1}, P. Bergeron\altaffilmark{2},
Daniel Eisenstein\altaffilmark{1}, H. C. Harris\altaffilmark{3},
S. J. Kleinman\altaffilmark{4}, Atsuko Nitta\altaffilmark{4}, Jurek
Krzesinski\altaffilmark{4,5} }

\altaffiltext{1}{Department of Astronomy and Steward Observatory,
University of Arizona, Tucson, AZ 85721}

\altaffiltext{2}{D\'epartement de Physique, Universit\'e de Montr\'eal,
C.P.~6128, Succ.~Centre-Ville, Montr\'eal, Qu\'ebec, Canada, H3C 3J7.}

\altaffiltext{3}{U.S. Naval Observatory, Flagstaff Station, Flagstaff AZ
86002-1149}

\altaffiltext{4}{Apache Point Observatory, P.O. Box 59, Sunspot NM
88349-0059}

\altaffiltext{5}{Mt. Suhora Observatory, Cracow Pedagogical
University, ul. Podchorazych 2, 30-084 Cracow, Poland}

\begin{abstract} 

We analyze the spectrum of an unusually-low mass white dwarf,
SDSSJ123410.37-022802.9 (0335-264-52000), found in our recent, white
dwarf catalog from the SDSS DR1 data release.  Two independent, model
atmosphere fits result in an accurate determination of atmospheric and
stellar parameters.  The more hands-on analysis yields $T_{\rm eff}$ =
17,470$\pm$750~K and $\log g$ = 6.38$\pm$0.05.  We argue that the object
cannot be a main sequence A star, a horizontal branch or subdwarf B
star.  Instead, it is interpreted as a very low mass white dwarf with a
core composed of helium, with mass ($\sim$0.18-0.19$M_\odot$), similar
to that published previously for the unusual companion to the
millisecond pulsar J1012+5307.  The star probably remains in a binary,
perhaps even with an undiscovered or dead pulsar companion.  However,
the companion might be a more-ordinary star, provided Roche lobe mass
transfer began shortly after the now-visible component left the main
sequence.  A second SDSS low mass white dwarf candidate is also
analyzed, but the spectrum for this fainter object is of poorer quality.
The sample appears to include additional, similar candidates, worthy of
more accurate observation.  Correct identification of any additional 
white dwarfs of extremely low mass requires careful observation and
interpretation. 

\end{abstract} 


\section{Introduction}

     Kleinman et al. (2004, hereafter K04) have released a catalog of
2561 white dwarf stars found in the first data release of the Sloan
Digital Sky Survey (Abazajian et al. 2003, DR1; see also
http://www.sdss.org/dr1).  As value added to the release, an automatic
fitting procedure (autofit) was used to estimate the parameters $T_{\rm
eff}$ and $\log~g$ for each white dwarf classified DA and DB.  Extensive
human checks of the ``outliers'' in the $T_{\rm eff}$ vs $\log~g$ plot
resulted in elimination of many invalid fits, due usually to a corrupted
spectrum or flux calibration.  However, K04 noted several white dwarfs
of unusually low surface gravity which appear to be valid, if usually
inaccurate, fits.

     While it has been known for a long time that the mass distribution
of field DA white dwarfs peaks near 0.6~$M_\odot$ (Koester, Schulz, \&
Weidemann 1980), the more precise determinations of $T_{\rm eff}$ and
$\log g$ made possible by accurate digital spectra and model atmosphere
fits to the Balmer lines demonstrated the reality of both separate high
mass and low mass components to this mass distribution (Bergeron,
Saffer, \& Liebert 1992; Bragaglia, Renzini, \& Bergeron 1995).  With
derived masses $\lesssim$0.5~$M_\odot$, the latter were inferred to have
helium cores, requiring envelope mass loss to a companion before the
progenitor reached the tip of the red giant branch and could ignite the
helium.  This prediction was dramatically confirmed by Marsh, Dhillon,
\& Duck (1995) and Maxted \& Marsh (1999), who found that a large
fraction of these stars and other low gravity DA white dwarfs were
indeed binaries with orbital periods of hours to days.  The inferred
companion was generally another white dwarf, sometimes detectable in the
spectrum.  In a few cases, an infrared excess was discovered, which
implied a low mass nondegenerate companion (Zuckerman \& Becklin 1992).

     The apparent helium white dwarfs in the Bergeron et al. (1992) and
Bragaglia et al. (1995) samples of stars with $T_{\rm eff}$ $<$ 50,000~K
had $\log~g$ as low as 7.22 and masses as low as 0.31~$M_\odot$
(PG~1241-010).  Very hot, low mass white dwarfs have radii much larger
than the final ``zero-temperature'' values.  For instance, in Bergeron
et al.'s (1994) analysis of hot DAO stars, the lowest gravity fitted was
for the star HZ34 at $\log~g$ = 6.61 and $T_{\rm eff}$ = 60,700~K.
Probably the lowest mass white dwarf reported up to now has a most
unusual evolutionary history.  It is the companion to the millisecond
pulsar J1012+5307.  Presumably it spun-up the pulsar during the prior
common envelope phase.  Van Kerkwijk, Bergeron, \& Kulkarni (1996) found
for the white dwarf $T_{\rm eff}$ = 8,550$\pm$25~K and $\log g$ =
6.75$\pm$0.07, from which they estimated a tiny mass of only 
0.16$\pm$0.02$M_\odot$.  Callanan, Garnavich, \& Koester (1998) derived
$T_{\rm eff}$ = 8670~K, $\log~g$ = 6.34, but the same mass of
0.16$\pm$0.02$M_\odot$ in an independent analysis.  

     Several stars in the K04 DA sample fit surface gravities lower than
those cited above.  For most of these mentioned in K04, Section 7.1, the
spectrophotometry of these stars near 20th magnitude is too poor for
accurate determinations of parameters.  Improved spectra would be
recommended.  For convenience, the SDSS photometry for the stars
mentioned in the cited paper is listed in Table~1.

     The main purpose of this paper is to develop the case for an 
extremely low mass for the best-observed example,
SDSSJ~123410.37-022802.9 (hereafter, SDSSJ1234), a result in which we
have high confidence.  We do include analysis of one other low mass
candidate, to illustrate what happens with a poorly-detected spectrum.
In Section 2, independent fits to the SDSS spectrophotometry of the
Balmer lines are presented.  In Section 3, we discuss the implications.

\section{Determination of parameters}

     SDSSJ1234 is a fairly bright star compared to most cataloged in
K04, at g = 17.87.  The proper motion measurement $\mu$ = 13
milliarcseconds per year is probably real.  The ``autofit'' techniques
are discussed in the cited paper.  For this star the fitted $T_{\rm
eff}$ = 17,500~K, with $\log~g$ = 6.375.  The fit is shown in Figure~13
of K04.
  
     Extraordinary claims, however, should be backed up by careful
analysis.  We therefore performed a careful, hands-on fit to the
spectrum using the independent pure hydrogen models of one of us (PB).
The method of analysis is that discussed in the previously-cited
Bergeron et al. (1992), Bragaglia et al. (1995), and many other Bergeron
papers.  The fit is shown in Figure~1(left).  The resulting parameters
are $T_{\rm eff}$ = 17,470~K$\pm$750, $\log~g$ = 6.38$\pm$0.05, in
excellent agreement with the autofit values in K04.  We can thus be
confident that these are good estimates of the parameters of this
unusual star.

     It is possible that other DA white dwarfs of unusually-low mass are
included in the DR1 Catalog. Several are mentioned in K04.  The problem
of analyzing a 2.5-meter spectrum of a much fainter candidate is
illustrated in Figure~1(right), an attempt to fit the Balmer lines of
SDSSJ~105611.03+653631.5 (g= 19.77) with Bergeron models.  The derived
parameters are $T_{\rm eff}$ = 21,910$\pm$1900~K, $\log~g$ =
7.07$\pm$0.09.  The autofit in the catalog gives 18,149$\pm$288~K and
6.73$\pm$0.10, respectively.  Reobservation of this star to achieve a
high signal-to-noise ratio is desirable to determine more accurate
parameters.

     We do not judge it to be worthwhile to show fits to the remaining
low mass candidates in the paper cited above.  For instance, the very
noisy spectrum of SDSSJ234536.48-010204.8 -- which is discussed more
extensively in K04 -- fit parameters of $T_{\rm eff}$ = 27,141~K,
$\log~g$ = 6.06 with the autofit package.  With Bergeron models and
techniques, the parameters are considerably different at 34,042~K and
6.77.  Clearly this is an example of a candidate which needs much more
precise observation. 

\section{The Nature of SDSSJ1234} 

We note that the SDSS colors of SDSSJ1234, $u-g$ = +0.32, $g-r$ = -0.35,
are similar to DA or subdwarf B stars of similar $T_{\rm eff}$.
Actually, these accurate colors place the object to the right of the
Bergeron $\log~g$=7 curve for hydrogen-atmosphere white dwarfs plotted
in the $u-g$-$g-r$ diagram (Figure~1) in the Harris et al. (2003) paper.
This paper presented the first sampling of white dwarfs from the SDSS.
The colors thus are in agreement with the determination that SDSSJ1234
has $\log~g$ less than 7.

Clearly, these colors are inconsistent with main sequence A or
horizontal branch A stars which have somewhat similar H line profiles,
but much larger Balmer jumps ($u-g\sim$1).  The fitted gravity is also
higher than for these types of stars, although this may not be obvious
from simple inspection of the hydrogen line spectrum.

Could SDSSJ1234 be a subdwarf B (sdB) star on the extended horizontal
branch?  Such stars can have similar colors.  However, this
interpretation cannot be correct.  Saffer et al. (1994) analyzed a
sample of bright, field sdB stars using a similar analysis of Balmer
line spectra at these lower surface gravities.  In their Figure 5, the
distribution is plotted in log~g vs. $T_{\rm eff}$.  Along this EHB, the
$\log~g$ increases with increasing Teff, but the upper limit to $\log~g$
at a given Teff is believed to be specified by models for the "zero age"
horizontal branch (HB), the location of the star at the start of the
core-helium burning phase.  Stars evolve upwards in luminosity to lower
$\log~g$ values as the He-burning proceeds.  HB models were also
plotted in the Saffer et al. figure.  It is shown that a star of
17,000~K is expected to have $\log~g$ no higher than 5.0, lower 
by more than an order of magnitude.

If the proper motion is real, the implied tangential velocity of the
star at the required absolute magnitude and distance for the horizontal
branch might leave it unbound to the galaxy.  At M$_V \sim$+2, g = 17.88
is roughly V = 17.98, then d = 15,700 pc.  Thus v$_{tan}$ = 4.738$\mu$d
$\sim$ 970~km~s$^{-1}$.  If instead we use the fitted M$_V$ = +8.22,
v$_{tan}$ = 55~km~s$^{-1}$, appropriate for the old or thick disk
population at a distance of about 900 pc.

The star lies in luminosity between the sdB/HB sequence and the sequence
of normal-mass white dwarfs.  At M$_V$ = +8.22, it is nearly six
magnitudes fainter than the HB at 17,000~K, and 2.7 magnitudes brighter
than a 17,000~K white dwarf with mass 0.6$M_\odot$.  The only possible
interpretation appears to be that it is a helium white dwarf with
unusually low mass and large radius.  The estimated radius is
0.047$R_\odot$, about four times that of a 0.6$M_\odot$ white dwarf.  At
an estimated M$_{bol}$ = +6.57, it is about one fifth of the luminosity
of the Sun!

How extreme are the parameters?  In Figure 2, we plot the six
evolutionary sequences of Althaus et al. (2001), and Serenelli et
al. (2001) in a temperature - gravity diagram, along with the positions
of the two SDSS stars analyzed in the previous section, and the white
dwarf companion to the millisecond pulsar discussed in the introduction.
SDSSJ1234 appears to fit a mass not significantly different from
0.196$M_\odot$, and essentially the same as these evolutionary sequences
would estimate for the pulsar companion.  However, these values also
depend on whether gravitational diffusion and a pure-hydrogen
atmospheric composition are correct assumptions, and on the assumed
thickness of the hydrogen envelope.  Note that Althaus et al. (2001)
estimated 0.169$M_\odot$ for the pulsar, using the lower gravity of
Callanan et al. (1998).  The values for these two unusual white dwarfs
obviously remain uncertain.  Finally, however, we note the even the mass
inferred for SDSSJ~105611.03+653631.5 is near the 0.31$M_\odot$ value
for the lowest mass white dwarf previously determined in the literature
(see the Introduction), apart from the pulsar companion.

It is very likely that SDSSJ1234 has a binary companion, as apparently
do the other low mass, helium-core white dwarfs.  If unusual evolution
in a common envelope with a neutron star is required for so much of the
original hydrogen envelope to be removed, then we would infer that this
white dwarf could have an undiscovered or dead pulsar companion.

If the binary separation is appropriate, however, it is possible that
the companion might just be a cool white dwarf following more ordinary
common envelope (CE) evolution.  For instance, if the progenitor of the
observed component were 1$M_\odot$, it leaves the main sequence and
reaches the so-called Sch\"onberg-Chandrasekhar limit with a helium core
of about 0.1$M_\odot$ (see, e.g. Kippenhahn \& Weigert 1990, p. 285).
The radius slowly expands from this point and the companion must be
located such that the evolving subgiant overfills its Roche lobe soon
afterwards, and begins unstable mass transfer to the degenerate
companion.  If the subsequent CE phase is sufficiently unstable, most of
the envelope mass may be lost quickly from the system, so that the core
mass of the evolving star does not grow past 0.16-0.19$M_\odot$.  Thus,
the phenomenon of ultra-low mass He white dwarfs may occur in the
ordinary field population from more-ordinary binary star evolution.

     We conclude by emphasizing that the correct identification of any
extremely low mass white dwarfs like SDSSJ1234 in various surveys
requires careful observation and interpretation.  In the past, some
might have been dismissed by inspection of the spectra as some other
class of star than a white dwarf. The SDSS will be an ideal sample for a
robust determination of frequency of such objects.

\acknowledgments 

This work was supported by the National Science Foundation through grant
AST-0307321 for study of white dwarfs found in the SDSS.  This work was
also supported in part by the NSERC Canada and by the Fund NATEQ
(Qu\'ebec).

\clearpage

\begin{figure} 

\caption{The simultaneous fits to the normalized Balmer lines of
SDSSJ1234 (left frame), and SDSSJ105611.03+653631.5 (right frame), using
line profiles calculated from Bergeron models.  This technique is
described in Bergeron papers given in the text. }

\end{figure} 

\begin{figure} 

\caption{The positions of the two SDSS stars analyzed here, and the
white dwarf companion to the millisecond pulsar J1012+5307, in a $T_{\rm
eff}$ vs. $\log g$ diagram, using the Bergeron fits.  Van Kerkwijk et
al. (1996) parameters are adopted here for the pulsar companion.
Several evolutionary tracks for very low mass helium core models from
Althaus et al. (2001) and Serenelli et al. (2001) are shown for
comparison. }

\end{figure} 

\begin{deluxetable}{lrrrrrrrrrr}

\tablenum{1}
\tablecaption{SDSS Photometry of Helium White Dwarf Candidates} 
\tablewidth{0pt}
\tablehead{
\colhead{SDSSJ} & \colhead{g} & \colhead{u-g} & \colhead{g-r} &
\colhead{r-i} & \colhead{i-z} & \colhead{$\sigma_u$} & \colhead{$\sigma_g$}
& \colhead{$\sigma_r$} & \colhead{$\sigma_i$} & \colhead{$\sigma_z$} } 

\startdata

002207.65--101423.5 & 19.77 & -0.11 & -0.30 & -0.19 & -0.29 & 0.05 &
0.02 & 0.03 & 0.04 & 0.22 \\
081136.34+461156.4 & 19.76 & -0.17 & -0.44 & -0.40 & 0.02 & 0.03 & 0.02
& 0.03 & 0.05 & 0.17 \\ 
102228.02+020035.2 & 20.15 & 0.20 & -0.30 & -0.36 & -0.13 & 0.05 & 0.04
& 0.04 & 0.07 & 0.28 \\ 
105611.03+653631.5 & 19.77 & 0.15 & -0.39 & -0.24 & -0.17 & 0.05 & 0.03
& 0.03 & 0.05 & 0.19 \\  
123410.37--022802.9 & 17.88 & 0.33 & -0.37 & -0.21 & -0.25 & 0.09 &
0.02 & 0.02 & 0.03 & 0.06 \\
130422.65+012214.2 & 19.33 & 0.45 & 0.19 & -0.02 & 0.04 & 0.04 & 0.03 &
0.02 & 0.02 & 0.06 \\ 
142601.48+010000.2 & 19.35 & 0.31 & -0.33 & -0.24 & -0.22 & 0.04 & 0.02
& 0.03 & 0.04 & 0.12 \\ 
234536.48--010204.8 & 19.50 & -0.28 & -0.42 & -0.28 & -0.32 & 0.03 &
0.02 & 0.02 & 0.04 & 0.16 \\
\enddata
\end{deluxetable}

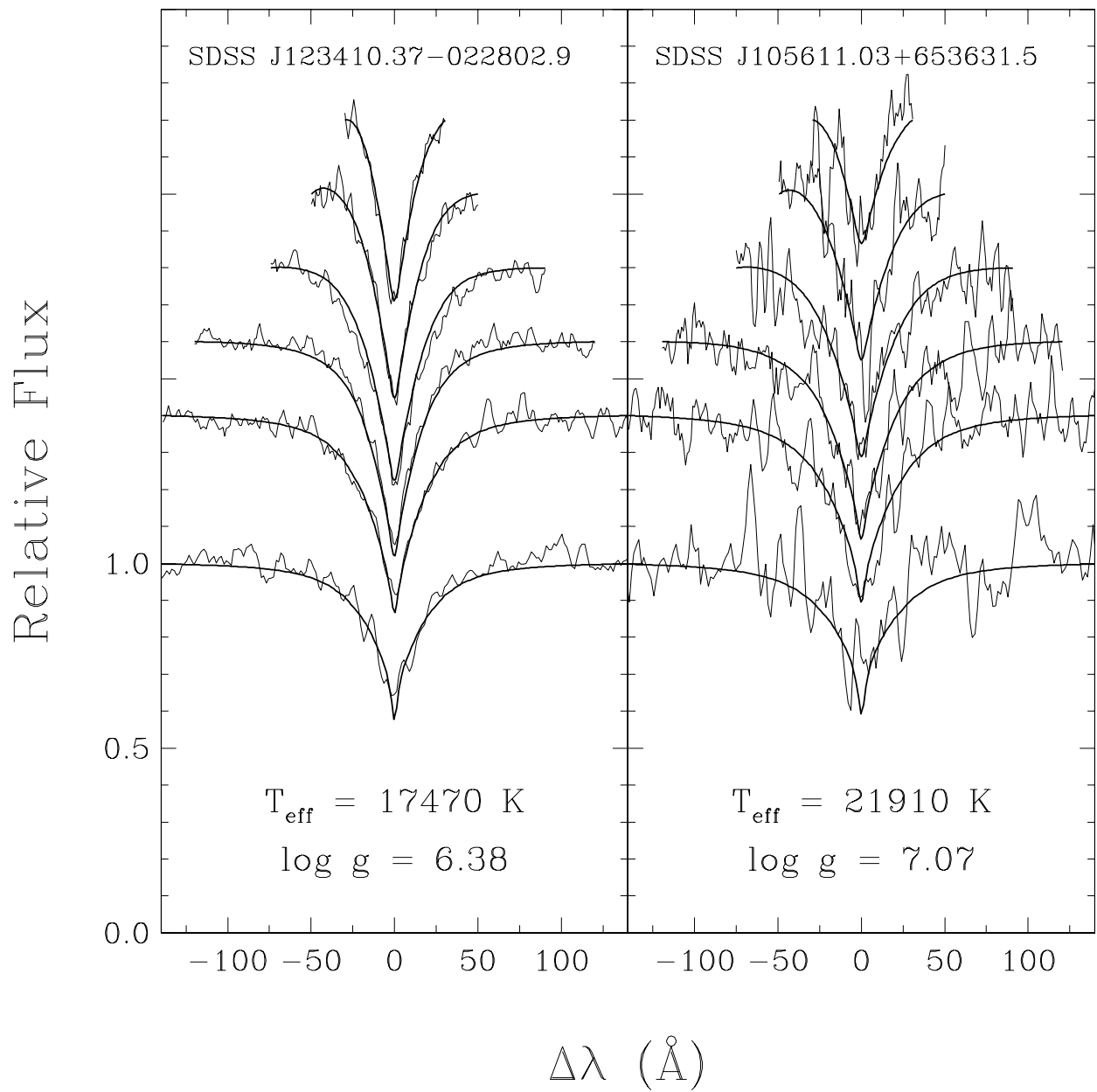
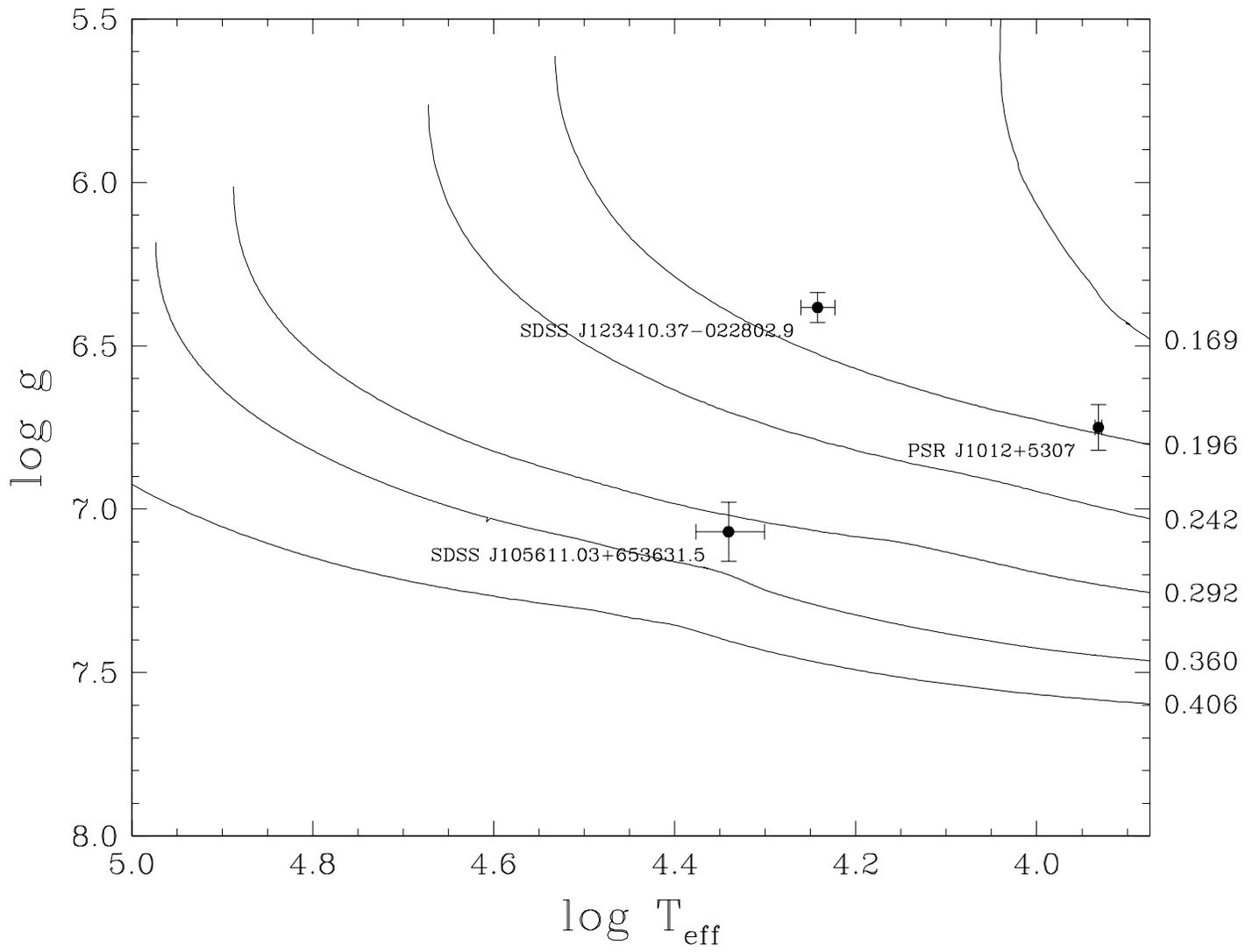
\end{document}